\documentclass[english,showpacs,preprint,superscriptaddress]{revtex4-1}
\usepackage{lmodern}
\usepackage{lmodern}
\usepackage[T1]{fontenc}
\usepackage[latin9]{inputenc}
\usepackage{amsmath}
\usepackage{amssymb}
\usepackage{esint}

\makeatletter

\usepackage{amsthm}\usepackage{latexsym}\usepackage{bm}\usepackage{amsfonts}

\usepackage{babel}

\usepackage{babel}
\usepackage{hyperref}

\makeatother

\usepackage{babel}
\begin{document}

\title{Geometric field theory and weak Euler-Lagrange equation for classical
relativistic particle-field systems}

\author{Peifeng Fan}

\affiliation{Institute of Plasma Physics, Chinese Academy of Sciences, Hefei,
Anhui 230031, China }

\affiliation{Department of Modern Physics, University of Science and Technology
of China, Hefei, Anhui 230026, China}

\author{Hong Qin }
\email{corresponding author: hongqin@ustc.edu.cn}

\affiliation{Department of Modern Physics, University of Science and Technology
of China, Hefei, Anhui 230026, China}

\affiliation{Center for Magnetic Fusion Theory, Chinese Academy of Sciences, Hefei
230031, China}

\affiliation{Plasma Physics Laboratory, Princeton University, Princeton, NJ 08543,
USA}

\author{Jian Liu}

\affiliation{Department of Modern Physics, University of Science and Technology
of China, Hefei, Anhui 230026, China}

\author{Nong Xiang}

\affiliation{Institute of Plasma Physics, Chinese Academy of Sciences, Hefei,
Anhui 230031, China }

\affiliation{Center for Magnetic Fusion Theory, Chinese Academy of Sciences, Hefei
230031, China}

\author{Zhi Yu}

\affiliation{Institute of Plasma Physics, Chinese Academy of Sciences, Hefei,
Anhui 230031, China }

\affiliation{Center for Magnetic Fusion Theory, Chinese Academy of Sciences, Hefei
230031, China}
\begin{abstract}
A manifestly covariant, or geometric, field theory for relativistic
classical particle-field system is developed. The connection between
space-time symmetry and energy-momentum conservation laws for the
system is established geometrically without splitting the space and
time coordinates, i.e., space-time is treated as one identity without
choosing a coordinate system. To achieve this goal, we need to overcome
two difficulties. The first difficulty arises from the fact that particles
and field reside on different manifolds. As a result, the geometric
Lagrangian density of the system is a function of the 4-potential
of electromagnetic fields and also a functional of particles' world-lines.
The other difficulty associated with the geometric setting is due
to the mass-shell condition. The standard Euler-Lagrange (EL) equation
for a particle is generalized into the geometric EL equation when
the mass-shell condition is imposed. For the particle-field system,
the geometric EL equation is further generalized into a weak geometric
EL equation for particles. With the EL equation for field and the
geometric weak EL equation for particles, symmetries and conservation
laws can be established geometrically. A geometric expression for
the energy-momentum tensor for particles is derived for the first
time, which recovers the non-geometric form in the existing literature
for a chosen coordinate system.
\end{abstract}

\pacs{52.35.Hr, 52.35.-g, 52.35.We, 42.50.Tx, 52.50.Sw }
\maketitle

\section{Introduction}

Energy-momentum conservation is a fundamental law of physics. It applies
to both quantum systems and classical systems. From the field-theoretical
point of view, energy-momentum conservation is fundamentally due to
the space-time symmetry of the Lagrangian (or Lagrangian density)
that the system admits \cite{Noether1918,Olver1993,Markakis2017Conservation}.
For example, the Lagrangian density for the electromagnetic field
is 
\begin{equation}
\mathcal{L}_{F}=\frac{1}{8\pi}\left[\left(-\frac{1}{c}\frac{\partial\mathbf{A}}{\partial t}-\nabla\varphi\right)^{2}-\left(\nabla\times\mathbf{A}\right)^{2}\right].\label{eq:17-1-1}
\end{equation}
The energy and momentum conservation laws of the system,

\begin{equation}
\frac{\partial}{\partial t}\left[\frac{\mathbf{E^{2}}+\mathbf{B^{2}}}{8\pi}\right]+\nabla\cdot\left[\frac{c}{4\pi}\mathbf{E}\times\mathbf{B}\right]=0,\label{eq:01}
\end{equation}
\begin{equation}
\frac{\partial}{\partial t}\left[\frac{\mathbf{E}\times\mathbf{B}}{4\pi c}\right]+\nabla\cdot\left[\frac{\mathbf{E^{2}}+\mathbf{B^{2}}}{8\pi}\mathbf{I}-\frac{\mathbf{EE}+\mathbf{BB}}{4\pi}\right]=0,\label{eq:02}
\end{equation}
can be derived from EL equations,

\begin{equation}
\frac{\partial\mathcal{L}_{F}}{\partial\varphi}-\frac{D}{D\boldsymbol{x}}\cdot\frac{\partial\mathcal{L}_{F}}{\partial\dot{\varphi}}=0,\label{eq:03-1}
\end{equation}
\begin{equation}
\frac{\partial\mathcal{L}_{F}}{\partial\mathbf{A}}-\frac{D}{D\boldsymbol{x}}\cdot\frac{\partial\mathcal{L}_{F}}{\partial\nabla\mathbf{A}}-\frac{D}{Dt}\frac{\partial\mathcal{L}_{F}}{\partial\mathbf{A}_{,t}}=0,\label{eq:04-1}
\end{equation}
and the symmetry conditions, 

\begin{equation}
\frac{\partial\mathcal{L}_{F}}{\partial t}=0,\label{eq:05-1}
\end{equation}
\begin{equation}
\frac{\partial\mathcal{L}_{F}}{\partial\boldsymbol{x}}=0.\label{eq:06-1}
\end{equation}
Here $D/D\boldsymbol{x}$ and $D/Dt$ denote partial derivative with
respect to $\boldsymbol{x}$ and $t$, respectively, when they operate
on a field defined on the space time. 

Geometrically, the symmetries of time and space of the Lagrangian
density are components of the space-time symmetry, 
\begin{equation}
\frac{\partial\mathcal{L}_{F}}{\partial\chi^{\mu}}=0,\quad(\mu=0,1,2,3),\label{eq:07-1}
\end{equation}
where $\chi^{\mu}$ is an arbitrary world point in 4 dimensional Minkowski
space-time, i.e., $\chi^{0}\equiv t$ and $\chi^{i}\equiv x^{i}$.
And the Lagrangian density in Eq.$\,$(\ref{eq:07-1}) can be equivalently
written as

\begin{equation}
\mathcal{L}_{F}=-\frac{1}{4\pi}\partial^{[\mu}A^{\nu]}\partial_{\mu}A_{\nu},\label{eq:07-2}
\end{equation}
where the 4-potential $A_{\nu}=\left(\varphi,\mathbf{-A}\right)$
is defined on the space-time. In this paper, we assume that the space-time
is endowed with a Lorentzian metric and the signature of the metric
is $\left(+---\right)$. Equation (\ref{eq:07-1}) is a manifestly
covariant form of Eqs.$\thinspace$(\ref{eq:05-1}) and (\ref{eq:06-1}).
In this paper, the phrase ``manifestly covariant'' will be substituted
by ``geometric'', which indicates that covariance is the intrinsic
coordinate-independent property of the physical system \cite{Qin2007}. 

Similarly, the EL equations (\ref{eq:03-1}) and (\ref{eq:04-1})
or the Maxwell equations can be written in the geometric form as

\begin{equation}
\frac{\partial\mathcal{L}_{F}}{\partial A_{\mu}}-\frac{D}{D\chi^{\nu}}\left[\frac{\partial\mathcal{L}_{F}}{\partial\left(\partial_{\nu}A_{\mu}\right)}\right]=0.\label{eq:08-1}
\end{equation}
Here, the operator $D/D\chi^{\nu}$ denotes partial derivative with
respect to $\chi^{\nu}$ while keeping $\chi^{\mu}(\mu\neq\nu)$ fixed,
when it is operated on a field on space-time. The geometric energy-momentum
conservation law is 
\begin{equation}
\partial_{\nu}T_{F}^{\mu\nu}=0,\quad(\mu,\nu=0,1,2,3),\label{eq:03}
\end{equation}
where $T_{F}^{\mu\nu}$ is the energy-momentum tensor of electromagnetic
fields and could be written in an explicit form as 
\begin{equation}
T_{F}^{\mu\nu}=\frac{1}{4\pi}\left(-F^{\text{\ensuremath{\mu\sigma}}}F_{\;\sigma}^{\nu}+\frac{1}{4\pi}\eta^{\mu\nu}F_{\sigma\rho}F^{\sigma\rho}\right),\quad(\mu,\nu,\sigma,\rho=0,1,2,3)\label{eq:04}
\end{equation}
where $F$ is electromagnetic tensor and $\eta$ is the Lorentzian
metric.  To briefly summarize, Eq. (\ref{eq:07-2}) is the geometric
form of Eq. (\ref{eq:17-1-1}), Eqs. (\ref{eq:03}) is that of Eqs.
(\ref{eq:01}) and (\ref{eq:02}). Similarly, Eq. (\ref{eq:08-1})
is the geometric form of Eqs. (\ref{eq:03-1}) and (\ref{eq:04-1}),
and Eq. (\ref{eq:07-1}) is that of Eqs. (\ref{eq:05-1}) and (\ref{eq:06-1}).
These, of course, are well-known \cite{Landau1975}. 

Classical particle-field systems, where many charged particles evolve
under the electromagnetic field generated self-consistently by the
particles, are often encountered in astrophysics, accelerator physics,
and plasma physics \cite{Brennan2014,Carrasco2016}. For these systems,
the relations between symmetries and conservation laws have only been
established recently by Qin et al \cite{Qin2014}. It turns out that
the standard EL equation for particles doesn't hold anymore, because
the dynamics of the particles and fields are defined on different
manifolds which have different dimensions. The electromagnetic fields
are defined on the space-time domain, whereas the particle trajectories
as fields are only defined on the time-axis. For particles, a weak
EL equation is established to replace the standard EL equation. It
was discovered that the weak EL equation can also link symmetries
with conservation laws as in standard field theory. This field theory
is non-relativistic, but it can be easily extended to relativistic
cases, which will be given in Sec.$\thinspace$\ref{sec:Relativistic-particle-field-syst}.
However, this approach is based on the split form of space and time.
In another word, it is not geometric. 

In this paper, we will geometrically reformulate the field theory
for classical particle-field system established in Ref. \cite{Qin2014}.
A geometric weak EL equation will be derived, and the energy-momentum
conservation will be geometrically derived from the space-time symmetry.
To achieve this goal, we need to overcome two difficulties. The first
difficulty arises from the fact that particles and field reside on
different manifolds as noticed in Ref.\,\cite{Qin2014}. In the geometric
setting, this difference is more prominent. Particles' dynamics are
characterized by world-lines on the space-time. They are defined on
$R^{1}$ as a field valued in the space-time, and the domain of the
particle fields can be the proper time or any other parameterization
for the world-lines. The world-line of a particle is uniquely defined,
independent of how it is parameterized. The electromagnetic field,
on the other hand, are defined on the space-time. The geometric Lagrangian
density of the system will be a function of the 4-potential of electromagnetic
fields and also a functional of particles' world-lines. This is qualitatively
different from the standard field theory, where the Lagrangian density
is a local function of the fields. The other difficulty associated
with the geometric setting is due to the mass-shell condition, which
exists even for the geometric variational principle for a single particle
\cite{Infeld1957}. The standard EL equation will be generalized into
a geometric EL equation when the mass-shell condition is imposed.
For the particle-field system, the geometric EL equation is further
generalized into a weak geometric EL equation. 

We emphasize that it is of significant theoretical and practical value
to put the physical laws governing the classical particle-field system
into geometric forms. The compact geometric forms, or manifestly Lorentz
invariant forms, are especially suitable for analysis in statistical
mechanics for relativistic plasmas \cite{Carrasco2016,Hakim1967a,Hakim1967b,Gedalin1996,Hornig1997,Baral2000,Tian2009,Tian2010,DAvignon2015,YangSD2016}.
They also serve as the theoretical foundations for developing Lorentz
covariant algorithms \cite{Wang2016} for numerical simulations. In
quantum field theory, both geometric and non-geometric forms are used.
The path integral approach is based on the geometric form of the Lagrangian
density, whiles the canonical quantization method is not manifestly
covariant because the space and time dimensions are split. The path
integral form of the quantum electrodynamics has been recently adopted
to develop a kinetic theory for magnetized plasmas when both quantum
and relativistic effects are important \cite{Shi2016}. 

 We note that energy-momentum conservation laws are well-known results,
and can be found, for example, in Ref. \cite{Landau1975}. However,
in the existing literature these conservation laws are not derived
from the underpinning symmetries. Often one can establish a conservation
law without knowing the underpinning symmetry. As in the case studied
here, establishing the connection between a symmetry and a conservation
law sometimes can be a difficult but rewarding task. 

This paper is organized as follows. In Sec.$\,$\ref{sec:Relativistic-particle-field-syst},
we will discuss how to extend the work by Qin, Burby and Davidson
to the relativistic case in a non-geometric way. The geometric Lagrangian
for a single particle and the corresponding geometric EL equation
are discussed in Sec.$\,$\ref{sec:The-geometrical-Lagrangian}. In
Sec.$\,$\ref{sec:Geometrical-Lagrangian-density}, the geometric
Lagrangian density for particle-field system and its properties are
discussed. The geometric weak EL equation and the link it provides
between conservation laws and space-time symmetries are given in Sec.$\,$\ref{sec:Geometrical-weak-Euler-Lagrangia}. 

\section{Non-geometric field theory and weak Euler-Lagrange equation for relativistic
particle-field systems \label{sec:Relativistic-particle-field-syst}}

The classical relativistic particle-field system in flat space is
governed by the following Newton-Maxwell equations,
\begin{equation}
\frac{d}{dt}(\gamma_{sp}m_{s}\dot{\mathbf{X}}_{sp})=q_{s}(\mathbf{E+}\frac{1}{c}\dot{\mathbf{X}}_{sp}\times\mathbf{B}),\label{eq:11-1}
\end{equation}
\begin{equation}
\nabla\cdot\mathbf{E}=4\pi\sum_{s,p}q_{s}\delta(\mathbf{X}_{sp}-\boldsymbol{x}),\label{eq:12-1}
\end{equation}
\begin{equation}
\nabla\times\mathbf{B}=\frac{4\pi}{c}\sum_{s,p}q_{s}\dot{\mathbf{X}}_{sp}\delta(\mathbf{X}_{sp}-\boldsymbol{x})+\frac{1}{c}\frac{\partial\mathbf{E}}{\partial t},\label{eq:13-1}
\end{equation}
\begin{equation}
\nabla\times\mathbf{E}=-\frac{1}{c}\frac{\partial\mathbf{B}}{\partial t},\label{eq:14}
\end{equation}
\begin{equation}
\nabla\cdot\mathbf{B}=0,\label{eq:15-1}
\end{equation}
where $\gamma_{sp}\equiv1/\sqrt{1-\dot{\boldsymbol{X}}_{sp}^{2}/c^{2}}$
is the relativistic factor for $p$-th particle of $s$-species, $m_{s}$
is the mass for any particle of $s$-species, $\mathbf{X}_{sp}$ is
the trajectory, $\boldsymbol{x}$ is a point in the configuration
space, and $\delta\left(\mathbf{X}_{sp}-\boldsymbol{x}\right)$ is
the Dirac delta function on 3D configuration space.  The basic Newton-Maxwell
equations (\ref{eq:11-1})-(\ref{eq:13-1}) can be equivalently written
as the Vlasov-Maxwell equations 
\begin{equation}
\frac{\partial F_{s}}{\partial t}+\boldsymbol{v}\cdot\frac{\partial F_{s}}{\partial\boldsymbol{x}}+q_{s}(\mathbf{E+}\frac{\boldsymbol{v}}{c}\times\mathbf{B})\cdot\frac{\partial F_{s}}{\partial\boldsymbol{p}}=0
\end{equation}
\begin{equation}
\nabla\cdot\mathbf{E}=4\pi\sum_{s}q_{s}\int F_{s}d^{3}\boldsymbol{p},
\end{equation}
\begin{equation}
\nabla\times\mathbf{B}=\frac{4\pi}{c}\sum_{s}q_{s}\int F_{s}\boldsymbol{v}d^{3}\boldsymbol{p}+\frac{1}{c}\frac{\partial\mathbf{E}}{\partial t},
\end{equation}
by defining the Klimontovich distribution function in phase space
as 
\begin{equation}
F_{s}=\sum_{p}\delta(\mathbf{X}_{sp}-\boldsymbol{x})\delta(\boldsymbol{p}_{sp}-\boldsymbol{p}).
\end{equation}
Here, $\boldsymbol{p}_{sp}=\gamma_{sp}m_{s}\dot{\mathbf{X}}_{sp}$
is the relativistic momentum of sp-particle.  The electric field $\mathbf{E}(t,\boldsymbol{x})$
and magnetic field $\mathbf{B}(t,\boldsymbol{x})$ are functions of
space and time. For this system, the action $\mathcal{A}$ is 

\begin{equation}
\mathcal{A}[\varphi,\mathbf{A},\mathbf{X}_{sp}]=\int\mathcal{L}_{PF}dtd^{3}x,\label{eq:16-1}
\end{equation}
where
\begin{eqnarray}
\mathcal{L}_{PF} & = & -\sum_{s,p}\gamma_{sp}^{-1}m_{s}c^{2}\delta(\mathbf{X}_{sp}-\boldsymbol{x})+\frac{q_{s}}{c}\mathbf{A}\cdot\dot{\mathbf{X}}_{sp}\delta(\mathbf{X}_{sp}-\boldsymbol{x})-q_{s}\varphi\delta(\mathbf{X}_{sp}-\boldsymbol{x})\nonumber \\
 & + & \frac{1}{8\pi}\left[\left(-\frac{1}{c}\frac{\partial\mathbf{A}}{\partial t}-\nabla\varphi\right)^{2}-\nabla\times\boldsymbol{A}\right]\label{eq:17-1}
\end{eqnarray}
is the Lagrangian density of this system \cite{Landau1975}. The variation
of \textbf{$\mathcal{A}$} induced by $\delta\mathbf{X}_{sp}$, \textbf{$\delta\mathbf{A}$
}and $\delta\varphi$ is

\begin{equation}
\delta\mathcal{A}=\sum_{s,p}\int dt\delta\mathbf{X}_{sp}\cdot\int E_{\mathbf{X_{sp}}}(\mathcal{L}_{PF})d^{3}x+\int E_{\varphi}(\mathcal{L}_{PF})\delta\varphi dtd^{3}x+\int E_{\mathbf{A}}(\mathcal{L}_{PF})\cdot\delta\mathbf{A}dtd^{3}x,\label{eq:18-1}
\end{equation}
where 
\begin{equation}
E_{\mathbf{X_{sp}}}(\mathcal{L}_{PF})\equiv\frac{\partial\mathcal{L}_{PF}}{\partial\mathbf{X}_{sp}}-\frac{D}{Dt}\cdot\frac{\partial\mathcal{L}_{PF}}{\partial\dot{\mathbf{X}}_{sp}},\label{eq:19-1}
\end{equation}
\begin{equation}
E_{\varphi}(\mathcal{L}_{PF})\equiv\frac{\partial\mathcal{L}_{PF}}{\partial\varphi}-\frac{D}{D\boldsymbol{x}}\cdot\frac{\partial\mathcal{L}_{PF}}{\partial\dot{\varphi}},\label{eq:20-1}
\end{equation}
\begin{equation}
E_{\mathbf{A}}(\mathcal{L}_{PF})\equiv\frac{\partial\mathcal{L}_{PF}}{\partial\mathbf{A}}-\frac{D}{D\boldsymbol{x}}\cdot\frac{\partial\mathcal{L}_{PF}}{\partial\nabla\mathbf{A}}-\frac{D}{Dt}\frac{\partial\mathcal{L}_{PF}}{\partial\mathbf{A}_{,t}}.\label{eq:21-1}
\end{equation}
For $\delta\mathcal{A}=0$ , we have 
\begin{equation}
\int E_{\mathbf{X_{sp}}}(\mathcal{L}_{PF})d^{3}\boldsymbol{x}=0,\label{eq:22}
\end{equation}
\begin{equation}
E_{\varphi}(\mathcal{L}_{PF})=0,\label{eq:23-1}
\end{equation}
\begin{equation}
E_{\mathbf{A}}(\mathcal{L}_{PF})=0\label{eq:24-1}
\end{equation}
due to the arbitrarinesses of $\delta\mathbf{X_{sp}}$, $\delta\varphi$
and $\delta\mathbf{A}$ in Eq.$\thinspace$(\ref{eq:18-1}). Equation$\thinspace$(\ref{eq:22})
will be called sub-manifold EL equation because it is defined only
on the time-axis after the integrating over spatial variable. Moreover,
substituting Eq.$\,$(\ref{eq:17-1}) into Eqs.$\thinspace$(\ref{eq:19-1})-(\ref{eq:24-1}),
we will recover Eqs.$\thinspace$(\ref{eq:11-1})-(\ref{eq:15-1})
by defining $\mathbf{E}=-\partial\mathbf{A}/\partial t-\nabla\varphi$
and $\mathbf{B=}\nabla\times\mathbf{A}.$

For Eq.$\thinspace$(\ref{eq:22}), in general, we expect that $E_{\mathbf{X_{sp}}}(\mathcal{L}_{PF})\neq0$
although the integral of this term vanishes. We can derive an explicit
expression for $E_{\mathbf{X_{sp}}}(\mathcal{L}_{PF})$ as
\begin{equation}
E_{\mathbf{X_{sp}}}(\mathcal{L}_{PF})=\frac{\partial\mathcal{L}_{PF}}{\partial\mathbf{X}_{sp}}-\frac{D}{Dt}\cdot\frac{\partial\mathcal{L}_{PF}}{\partial\dot{\mathbf{X}}_{sp}}=\frac{\partial}{\partial x}\left(H_{sp}-\mathbf{P}_{sp}\cdot\dot{\mathbf{X}}_{sp}\right)+\frac{\partial}{\partial\boldsymbol{x}}\cdot\left(\dot{\mathbf{X}}_{sp}\mathbf{P}_{sp}\right),\label{eq:27}
\end{equation}
where 
\begin{equation}
H_{sp}=\left(\gamma_{sp}m_{s}c^{2}+q_{s}\varphi\right)\delta\left(\mathbf{X}_{sp}-\boldsymbol{x}\right),\label{eq:28-1}
\end{equation}
\begin{equation}
\mathbf{P}_{sp}=\left(\gamma_{sp}m_{s}\dot{\mathbf{X}}_{sp}+\frac{q_{s}}{c}\mathbf{A}\right)\delta\left(\mathbf{X}_{sp}-\boldsymbol{x}\right).\label{eq:29}
\end{equation}
Equation (\ref{eq:27}) is called weak EL equation \cite{Qin2014},
and the qualifier ``weak'' is used to indicate the fact that only
the spatial integral of $E_{\mathbf{X_{sp}}}(\mathcal{L}_{PF})$ is
zero (see Eq.$\thinspace$(\ref{eq:22})). 

Next, we define the symmetry of the action $\mathcal{A}[\varphi,\mathbf{A},\mathbf{X}_{sp}]$
to be a group of transformation
\begin{equation}
\left(t,\boldsymbol{x},\varphi,\mathbf{A},\mathbf{X}_{sp}\right)\longmapsto\left(\tilde{t},\tilde{\boldsymbol{x}},\tilde{\varphi},\tilde{\mathbf{A}},\tilde{\mathbf{X}}_{sp}\right),
\end{equation}
such that 
\begin{equation}
\int\mathcal{L}_{PF}\left(t,\boldsymbol{x},\varphi,\mathbf{A},\mathbf{X}_{sp}\right)dtd^{3}x=\int\tilde{\mathcal{L}}_{PF}\left(\tilde{t},\tilde{\boldsymbol{x}},\tilde{\varphi},\tilde{\mathbf{A}},\tilde{\mathbf{X}}_{sp}\right)dtd^{3}\boldsymbol{x}.\label{eq:31-1}
\end{equation}
For our Lagrangian density (see Eq.$\thinspace$(\ref{eq:17-1})),
if the group transformation is the time-translation, 
\begin{equation}
\left(\tilde{t},\tilde{\boldsymbol{x}},\tilde{\varphi},\tilde{\mathbf{A}},\tilde{\mathbf{X}}_{sp}\right)=\left(t+\epsilon,\boldsymbol{x},\varphi,\mathbf{A},\mathbf{X}_{sp}\right),\quad\epsilon\in\mathbb{R}\label{eq:32-1}
\end{equation}
the condition (\ref{eq:31-1}) will be satisfied because
\begin{equation}
\frac{\partial\mathcal{L}_{PF}}{\partial t}=0.\label{eq:33-1}
\end{equation}
Using weak EL equations (\ref{eq:27}) for particles, the EL equations
for fields (see Eqs.$\thinspace$(\ref{eq:20-1}), (\ref{eq:21-1}),
(\ref{eq:23-1}), (\ref{eq:24-1})), we obtain the energy conservation
law
\begin{equation}
\frac{\partial}{\partial t}\left[\sum_{s,p}\gamma_{sp}m_{s}c^{2}\delta\left(\mathbf{X}_{sp}-\boldsymbol{x}\right)+\frac{\mathbf{E^{2}}+\mathbf{B^{2}}}{8\pi}\right]+\nabla\cdot\left[\sum_{s,p}\gamma_{sp}m_{s}c^{2}\dot{\mathbf{X}}_{sp}\delta\left(\mathbf{X}_{sp}-\boldsymbol{x}\right)+\frac{c}{4\pi}\mathbf{E}\times\mathbf{B}\right]=0.\label{eq:34-1}
\end{equation}
Equation$\thinspace$(\ref{eq:31-1}) holds for spatial-translation
as well
\begin{equation}
\left(\tilde{t},\tilde{\boldsymbol{x}},\tilde{\varphi},\tilde{\mathbf{A}},\tilde{\mathbf{X}}_{sp}\right)=\left(t,\boldsymbol{x}+\epsilon\mathbf{X},\varphi,\mathbf{A},\mathbf{X}_{sp}+\epsilon\mathbf{X}\right),\quad\epsilon\in\mathbb{R}\label{eq:35-1}
\end{equation}
because $\mathcal{L}_{PF}$ satisfies 

\begin{equation}
\frac{\partial\mathcal{L}_{PF}}{\partial\boldsymbol{x}}+\sum_{s,p}\frac{\partial\mathcal{L}_{PF}}{\partial\mathbf{X}_{sp}}=0.\label{eq:36-1}
\end{equation}
As a consequence, the momentum conservation law due to this symmetry
can be written as 

\begin{equation}
\frac{\partial}{\partial t}\left[\sum_{s,p}\gamma_{sp}m_{s}\dot{\mathbf{X}}_{sp}+\frac{\mathbf{E}\times\mathbf{B}}{4\pi c}\right]+\nabla\cdot\left[\sum_{s,p}\gamma_{sp}m_{s}\dot{\mathbf{X}}_{sp}\dot{\mathbf{X}}_{sp}+\frac{\mathbf{E^{2}}+\mathbf{B^{2}}}{8\pi}\mathbf{I}-\frac{\mathbf{EE}+\mathbf{BB}}{4\pi}\right]=0.\label{eq:37-1}
\end{equation}

Details of the derivation can be found in Ref.\,\cite{Qin2014}.
However, in relativistic cases, this split form of space and time
is not elegant. The space and time should be treated as one identity
in the most fundamental and geometric approach. In the following sections,
we will explore the geometric way to establish the relations between
symmetries and conservation laws. 

 Now, let\textquoteright s discuss the necessity of the weak Euler-Lagrange
equation introduced in the present study. We note that one can construct
a velocity field $\boldsymbol{v}(\boldsymbol{x},t)$ on spacetime
using particles\textquoteright{} trajectories $\mathbf{X}_{sp}\left(t\right)$
as $\boldsymbol{v}\left(\boldsymbol{x},t\right)=\Sigma_{s,p}\boldsymbol{v}_{sp}\left(\boldsymbol{x},t\right)\delta\left(\boldsymbol{x}-\mathbf{X}_{sp}\left(t\right)\right)$.
However, in the variation procedure, we cannot treat the velocity
field $\boldsymbol{v}\left(\boldsymbol{x},t\right)$ defined this
way on spacetime as an independent field that can be varied freely
by an arbitrary $\delta\boldsymbol{v}\left(\boldsymbol{x},t\right)$.
In the variation procedure, the quantities that can be independently
varied are the 4-potential $A\left(\boldsymbol{x},t\right)$ and particles\textquoteright{}
trajectories $\mathbf{X}_{sp}\left(t\right)$. Obviously, $A\left(\boldsymbol{x},t\right)$
and $\mathbf{X}_{sp}\left(t\right)$ are defined on different domain.
This is the reason that we need to introduce the weak Euler-Lagrange
equation to overcome this difficulty. 

Having said that, there is indeed an alternative approach, if we insist
on varying the velocity field $\boldsymbol{v}\left(\boldsymbol{x},t\right)$,
instead of $\mathbf{X}_{sp}\left(t\right)$. In this case, the velocity
field $\boldsymbol{v}\left(\boldsymbol{x},t\right)$ cannot be varied
arbitrarily. The variation $\delta\boldsymbol{v}\left(\boldsymbol{x},t\right)$
needs to satisfy certain constraints. A systematic approach for this
kind of constrained variation has been developed in the context of
Euler-Poincare reduction \cite{Holm1998,Squire2013,ZhouY2014}. The
dynamic equation resulting from the constrained variation will be
very different from the standard Euler-Lagrange equation, and will
complicate the symmetry analysis. In the present study, we will not
pursue along this route. Note that the standard Euler-Lagrange equation
needs to be amended in either approach.

\section{Geometric Lagrangian and geometric Euler-Lagrangian equation for
a single particle\label{sec:The-geometrical-Lagrangian}}

For a classical particle, the action can be expressed as 

\begin{equation}
\mathcal{A}=\int_{t_{1}}^{t_{2}}L(\boldsymbol{x},\dot{\boldsymbol{x}},t)dt,
\end{equation}
where \textbf{$L(\boldsymbol{x},\dot{\boldsymbol{x}},t)$} is the
Lagrangian. Apply the principle of least action,

\begin{equation}
0=\delta\mathcal{A}=\int_{t_{1}}^{t_{2}}\delta L(\boldsymbol{x},\dot{\boldsymbol{x}},t)dt=\int_{t_{1}}^{t_{2}}\left[\frac{\partial L}{\partial\boldsymbol{x}}-\frac{d}{dt}\left(\frac{\partial L}{\partial\dot{\boldsymbol{x}}}\right)\right]\cdot\delta\boldsymbol{x}dt+\left[\frac{\partial L}{\partial\dot{\boldsymbol{x}}}\cdot\delta\boldsymbol{x}\right]_{t_{1}}^{t_{2}}.\label{eq:=00003D00FF15}
\end{equation}
Because of the fact that $\delta\boldsymbol{x}(t_{1})=\delta\boldsymbol{x}(t_{2})=0$
and the arbitrariness of $\delta\boldsymbol{x}$, we obtain the EL
equation for the dynamics of the particle

\begin{equation}
\frac{\partial L}{\partial\boldsymbol{x}}-\frac{d}{dt}\left(\frac{\partial L}{\partial\dot{\boldsymbol{x}}}\right)=0.\label{eq:=00003D00FF16}
\end{equation}

For relativistic particles, the variational principle is more involved.
In special relativity, the Lagrangian should be Lorentz invariant.
For this reason, the integral variable of the action should be the
proper time $\tau$ and the action is 

\begin{equation}
\mathcal{A}=\int_{p_{1}}^{p_{2}}\widetilde{L}(\mathbf{\mathbf{\mathbf{\mathbf{\chi}}}},\dot{\chi},\tau)d\tau,\label{eq:=00003D00FF17}
\end{equation}
where $\mathbf{\mathbf{\chi}}$ and $\dot{\chi}\equiv d\mathbf{\chi}/d\tau$
are the world-line and 4-velocity of the particle, respectively. The
Lagrangian $\widetilde{L}$ is manifestly covariant\textbf{ }\cite{Infeld1957}\textbf{
}and is called geometric Lagrangian in the present study. The integral
in Eq.$\,$(\ref{eq:=00003D00FF17}) is along any possible world-lines
passing $p_{1}$ and $p_{2}$ in space-time. If we parameterize the
world-line by the proper time $\tau,$ then the Eq.$\,$(\ref{eq:=00003D00FF17})
can be written as

\begin{equation}
\mathcal{A}=\int_{\tau_{1}}^{\tau_{2}}\widetilde{L}(\mathbf{\mathbf{\mathbf{\mathbf{\chi}}}},\dot{\chi},\tau)d\tau,
\end{equation}
where $\tau_{1}$ is the proper time at the beginning and $\tau_{2}$
is the proper time at the end. We can choose the parameter $\tau_{1}$
to be same for any possible world-line. But, the parameter $\tau_{2}$
are generally different because of the different length for different
world-lines passing the given space-time point $p_{2}$, which means
$\delta(d\tau)\neq0$. This is different from the non-relativistic
case, which implicitly assumes $\delta(dt)=0$. The Hamiltonian principle
is 
\begin{equation}
0=\delta\mathcal{A}=\int_{\tau_{1}}^{\tau_{2}}\delta\widetilde{L}(\mathbf{\mathbf{\mathbf{\mathbf{\chi}}}},\dot{\chi},\tau)d\tau+\int_{\tau_{1}}^{\tau_{2}}\widetilde{L}(\mathbf{\mathbf{\mathbf{\mathbf{\chi}}}},\dot{\chi},\tau)\delta(d\tau),\label{eq:9}
\end{equation}
where 
\begin{eqnarray}
\delta(d\tau) & = & \frac{1}{c^{2}}\delta\left(\frac{d\chi^{\mu}d\chi_{\mu}}{d\tau}\right)=\frac{1}{c^{2}}\delta\left[d\chi^{\mu}\frac{d\chi_{\mu}}{d\tau}\right]=\frac{1}{c^{2}}\delta\left[d\chi^{\mu}\dot{\chi}_{\mu}\right]\nonumber \\
 & \text{=} & \frac{1}{c^{2}}\delta\left(d\chi^{\mu}\right)\dot{\chi}_{\mu}+\frac{1}{c^{2}}d\chi^{\mu}\delta\left(\dot{\chi}_{\mu}\right)=\frac{1}{c^{2}}\delta\left(d\chi^{\mu}\right)\dot{\chi}_{\mu}+\frac{1}{c^{2}}\dot{\chi}^{\mu}\delta\left(\dot{\chi}_{\mu}\right)d\tau.\label{eq:10}
\end{eqnarray}
Here, $\chi^{\mu}$ and $\dot{\chi}^{\mu}$ $\left(\mu=0,1,2,3\right)$
are the components of the world-line and 4-veolocity of the particle.
From the mass-shell condition 
\begin{equation}
P^{\mu}P_{\mu}=m_{0}^{2}c^{2}\mbox{ or \ensuremath{\dot{\chi}^{\mu}\dot{\chi}_{\mu}}=\ensuremath{c^{2}}},\label{eq:11}
\end{equation}
where $P^{\mu}=m_{0}\dot{\chi}^{\mu}$ are the components of 4-momentum
and $m_{0}$ is the rest mass. We have

\begin{equation}
0=\delta\left(c^{2}\right)=\delta\left(\dot{\chi}^{\mu}\dot{\chi}_{\mu}\right)=\delta\left(\dot{\chi}^{\mu}\right)\dot{\chi}_{\mu}+\dot{\chi}^{\mu}\delta\left(\dot{\chi}_{\mu}\right)=2\dot{\chi}^{\mu}\delta\left(\dot{\chi}_{\mu}\right),\label{eq:13}
\end{equation}
which means that Eq.$\,$(\ref{eq:10}) can be reduced to

\begin{equation}
\delta(d\tau)=\frac{1}{c^{2}}\dot{\chi}_{\mu}\delta\left(d\chi^{\mu}\right).
\end{equation}
The second term on the right-hand side of Eq.\,\eqref{eq:9} is 
\begin{eqnarray}
\frac{1}{c^{2}}\int_{\tau_{1}}^{\tau_{2}}\widetilde{L}\dot{\chi}_{\mu}\delta\left(d\chi^{\mu}\right) & = & \frac{1}{c^{2}}\int_{\tau_{1}}^{\tau_{2}}\widetilde{L}\dot{\chi}_{\mu}\frac{d\left(\delta\chi^{\mu}\right)}{d\tau}d\tau\nonumber \\
 & = & -\frac{1}{c^{2}}\int_{\tau_{1}}^{\tau_{2}}\frac{d}{d\tau}\left[\widetilde{L}\dot{\chi}_{\mu}\right]\left(\delta\chi^{\mu}\right)d\tau+\left[\widetilde{L}\dot{\chi}_{\mu}\left(\delta\chi^{\mu}\right)\right]_{\tau_{1}}^{\tau_{2}}\nonumber \\
 & = & -\frac{1}{c^{2}}\int_{\tau_{1}}^{\tau_{2}}\frac{d}{d\tau}\left[\widetilde{L}\dot{\chi}_{\mu}\right]\delta\chi^{\mu}d\tau.\label{eq:15}
\end{eqnarray}
The first term on the right-hand side of Eq.\,\eqref{eq:9} is

\begin{eqnarray}
 &  & \int_{\tau_{1}}^{\tau_{2}}\left[\frac{\partial\widetilde{L}}{\partial\chi^{\mu}}\delta\chi^{\mu}+\frac{\partial\widetilde{L}}{\partial\dot{\chi}^{\mu}}\delta\dot{\chi}^{\mu}\right]d\tau\nonumber \\
 &  & =\int_{\tau_{1}}^{\tau_{2}}\left\{ \frac{\partial\widetilde{L}}{\partial\chi^{\mu}}\delta\chi^{\mu}+\frac{\partial\widetilde{L}}{\partial\dot{\chi}^{\mu}}\left[\frac{d\left(\delta\chi^{\mu}\right)}{d\tau}-\frac{1}{c^{2}}\dot{\chi}^{\mu}\dot{\chi}_{\nu}\frac{d\left(\delta\chi^{\nu}\right)}{d\tau}\right]\right\} d\tau\nonumber \\
 &  & =\int_{\tau_{1}}^{\tau_{2}}\frac{\partial\widetilde{L}}{\partial\chi^{\mu}}\delta\chi^{\mu}d\tau+\int_{\tau_{1}}^{\tau_{2}}\left[\frac{\partial\widetilde{L}}{\partial\dot{\chi}^{\mu}}-\frac{1}{c^{2}}\frac{\partial\widetilde{L}}{\partial\dot{\chi}^{\nu}}\dot{\chi}^{\nu}\dot{\chi}_{\mu}\right]\frac{d\left(\delta\chi^{\mu}\right)}{d\tau}d\tau\nonumber \\
 &  & =\int_{\tau_{1}}^{\tau_{2}}\frac{\partial\widetilde{L}}{\partial\chi^{\mu}}\delta\chi^{\mu}d\tau-\int_{\tau_{1}}^{\tau_{2}}\frac{d}{d\tau}\left[\frac{\partial\widetilde{L}}{\partial\dot{\chi}^{\mu}}-\frac{1}{c^{2}}\dot{\chi}^{\nu}\frac{\partial\widetilde{L}}{\partial\dot{\chi}^{\nu}}\dot{\chi}_{\mu}\right]\left(\delta\chi^{\mu}\right)d\tau\nonumber \\
 &  & +\left\{ \left[\frac{\partial\widetilde{L}}{\partial\dot{\chi}^{\mu}}-\frac{1}{c^{2}}\dot{\chi}^{\nu}\frac{\partial\widetilde{L}}{\partial\dot{\chi}^{\nu}}\dot{\chi}_{\mu}\right]\delta\chi^{\mu}\right\} _{\tau_{1}}^{\tau_{2}}\nonumber \\
 &  & =\int_{\tau_{1}}^{\tau_{2}}\left[\frac{\partial\widetilde{L}}{\partial\chi^{\mu}}-\frac{d}{d\tau}\left(\frac{\partial\widetilde{L}}{\partial\dot{\chi}^{\mu}}-\frac{1}{c^{2}}\dot{\chi}^{\nu}\frac{\partial\widetilde{L}}{\partial\dot{\chi}^{\nu}}\dot{\chi}_{\mu}\right)\right]\delta\chi^{\mu}d\tau,\label{eq:18-2}
\end{eqnarray}
where used if made of the following identity,

\begin{equation}
\delta\dot{\chi}^{\mu}=\frac{d\left(\delta\chi^{\mu}\right)d\tau-\frac{1}{c^{2}}\dot{\chi}^{\nu}d\chi^{\mu}d\left(\delta\chi^{\nu}\right)}{\left(d\tau\right)^{2}}=\frac{d\left(\delta\chi^{\mu}\right)}{d\tau}-\frac{1}{c^{2}}\dot{\chi}^{\mu}\dot{\chi}_{\nu}\frac{d\left(\delta\chi^{\nu}\right)}{d\tau}.\label{eq:17}
\end{equation}
Substituting Eqs.$\,$(\ref{eq:15}) and (\ref{eq:18-2}) into Eq.$\,$(\ref{eq:9}),
we obtain

\[
0=\delta\mathcal{A}=\int_{\tau_{1}}^{\tau_{2}}\left\{ \frac{\partial\widetilde{L}}{\partial\chi^{\mu}}-\frac{d}{d\tau}\left[\frac{\partial\widetilde{L}}{\partial\dot{\chi}^{\mu}}-\frac{1}{c^{2}}\left(\dot{\chi}^{\nu}\frac{\partial\widetilde{L}}{\partial\dot{\chi}^{\nu}}-\widetilde{L}\right)\dot{\chi}_{\mu}\right]\right\} \delta\chi^{\mu}d\tau,
\]
which implies

\begin{equation}
\frac{\partial\widetilde{L}}{\partial\chi^{\mu}}-\frac{d}{d\tau}\left[\frac{\partial\widetilde{L}}{\partial\dot{\chi}^{\mu}}-\frac{1}{c^{2}}\left(\dot{\chi}^{\nu}\frac{\partial\widetilde{L}}{\partial\dot{\chi}^{\nu}}-\widetilde{L}\right)\dot{\chi}_{\mu}\right]=0,\label{eq:19}
\end{equation}
due to the arbitrariness of $\delta\chi^{\mu}$. Equation (\ref{eq:19})
will be called geometric EL equation. We note that it has been derived
by Infeld using the method of Lagrange multiplier \cite{Infeld1957}. 

\section{Geometric Lagrangian density\label{sec:Geometrical-Lagrangian-density}}

For particle-field systems, we need to find the density of the geometric
Lagrangian. We start from Eq.$\,$(\ref{eq:16-1}), the non-geometric
form of the action of our system. It can be written as \cite{Landau1975}

\begin{equation}
\mathcal{A}=-\sum_{s,p}\int m_{s}c^{2}d\tau-\sum_{s,p}\int\frac{q_{s}}{c}A_{\mu}d\chi_{sp}^{\mu}-\frac{1}{16\pi c}\int F_{\mu\nu}F^{\mu\nu}d\Omega,\label{eq:20}
\end{equation}
where $\chi_{sp}^{\mu}$ $(\mu=0,1,2,3)$ are components of the world-line
of $p$th particle of $s$-species, $\tau$ is the proper time, the
4-potential $A_{\mu}$ and the field-strength tensor $F^{\mu\nu}$
are the functions on the Minkowski space. The boundary of the domain
is taken to be at the infinity. 

Equation (\ref{eq:20}) can be easily translated into another form,

\begin{equation}
\mathcal{A}=\sum_{s,p}\int\left(-m_{s}c^{2}-\frac{q_{s}}{c}A_{\mu}\dot{\chi}^{\mu}(\tau)\right)d\tau-\frac{1}{4\pi c}\int\partial^{[\mu}A^{\nu]}\partial_{\mu}A_{\nu}d\Omega,\label{eq:21}
\end{equation}
by using the relations $\dot{\chi}^{\mu}=d\chi^{\mu}/d\tau$ and $F_{\mu\nu}=\partial_{\mu}A_{\nu}-\partial_{\nu}A_{\mu}$,
where the symbol $[\mu\;\nu]$ in Eq.$\,$(\ref{eq:21}) means the
anti-symmetrization of the indexes of $\mu$ and $\nu$. The action
is manifestly covariant or geometric. However, to obtain the Lagrangian
density of the particle-field systems, we should change it into another
form by multiplying the first term in Eq.$\,$(\ref{eq:21}) with
the equation

\begin{equation}
\int\delta\left(\chi_{sp}-\chi\right)d\Omega=1,\label{eq:iv-54}
\end{equation}
where $\chi$ is an arbitrary world point in Minkowski space, and
$\delta\left(\chi_{sp}-\chi\right)$ is the Dirac delta function.
Then equation (\ref{eq:21}) becomes

\begin{equation}
\mathcal{A}=\int\left[\sum_{s,p}\int\left(-m_{s}c^{2}-\frac{q_{s}}{c}A_{\mu}\dot{\chi}^{\mu}(\tau)\right)\delta\left(\chi_{sp}-\chi\right)d\tau\right]d\Omega-\frac{1}{4\pi c}\int\partial^{[\mu}A^{\nu]}\partial_{\mu}A_{\nu}d\Omega.\label{eq:23}
\end{equation}
The geometric Lagrangian density is easy to read off from Eq.$\,$(\ref{eq:23})
as

\begin{equation}
\mathcal{L}=\sum_{s,p}\int\left(-m_{s}c^{2}-\frac{q_{s}}{c}A_{\mu}\dot{\chi}_{sp}^{\mu}(\tau)\right)\delta\left(\chi_{sp}-\chi\right)d\tau-\frac{1}{4\pi c}\partial^{[\mu}A^{\nu]}\partial_{\mu}A_{\nu}.\label{eq:24}
\end{equation}
The geometric Lagrangian above can be simplified as

\begin{equation}
\mathcal{L}=\mathcal{L}_{P}+\mathcal{L}_{F}=\int\hat{\mathcal{L}_{P}}d\tau+\mathcal{L}_{F}\label{eq:25}
\end{equation}
by defining 
\begin{equation}
\hat{\mathcal{L}_{P}}=\sum_{s,p}\left(-m_{s}c^{2}-\frac{q_{s}}{c}A_{\mu}\dot{\chi}_{sp}^{\mu}(\tau)\right)\delta\left(\chi_{sp}-\chi\right),\label{eq:26}
\end{equation}
\begin{equation}
\mathcal{L}_{P}=\int\hat{\mathcal{L}_{P}}\left(\chi_{sp},\dot{\chi}_{sp},\chi,A\right)d\tau
\end{equation}
and
\begin{equation}
\mathcal{L}_{F}=-\frac{1}{4\pi c}\partial^{[\mu}A^{\nu]}\partial_{\mu}A_{\nu}.\label{eq:28}
\end{equation}

Here, $\hat{\mathcal{L}_{P}}$ is a function of $\chi$ , $\chi_{sp}$
, $\dot{\chi}_{sp}$ and $A$. On the other hand, $\mathcal{L}_{P}$
can be regarded as a functional of the world-lines of particles and
a function of the space-time and the field $A$. Note that $\mathcal{L}_{F}$
in Eq.$\,$(\ref{eq:28}) is different from $\mathcal{L}_{F}$ in
Eq.$\,$(\ref{eq:07-2}) by a constant $c$. This is caused by the
fact that in the geometric form of the action given by Eq.\,\eqref{eq:23},
the volume form $d\Omega$ has the dimension of $\text{[length]}^{4}.$

\section{Geometric weak Euler-Lagrangian equation and energy-momentum conservation\label{sec:Geometrical-weak-Euler-Lagrangia}}

Now we determine how the action given by Eq.$\,$(\ref{eq:23}) and
geometric Lagrangian density vary in response to the variation of
$\delta\chi_{sp}$ and $\delta A$. From Eqs.$\,$(\ref{eq:23})-(\ref{eq:28}),
the variation of action of the particle-field system can be written
as 
\begin{eqnarray}
\delta\mathcal{A} & = & \int\delta\left[\int\hat{\mathcal{L}_{P}}\left(\chi_{sp},\dot{\chi}_{sp},\chi,A\right)d\tau\right]_{A}d\Omega\nonumber \\
 & + & \int\delta\left[\int\hat{\mathcal{L}_{P}}\left(\chi_{sp},\dot{\chi}_{sp},\chi,A\right)d\tau\right]_{\chi_{sp},\dot{\chi}_{sp}}d\Omega+\int\delta\mathcal{L}_{F}d\Omega,\label{eq:30}
\end{eqnarray}
where the notation $\delta\left[\int\hat{\mathcal{L}}_{P}\left(\chi_{sp},\dot{\chi}_{sp},\chi,A\right)d\tau\right]_{\alpha}$
($\alpha=A,\chi_{sp},\dot{\chi}_{sp}$) means keeping $\alpha$ fixed.

For the first term on the right-hand side of Eq.\,\eqref{eq:30},
it can be treated by the same procedure for the derivation of Eq.\,\eqref{eq:19},
\begin{eqnarray}
 &  & \int\delta\left[\int\hat{\mathcal{L}_{P}}\left(\chi_{sp},\dot{\chi}_{sp},\chi,A\right)d\tau\right]_{A}d\Omega\nonumber \\
 &  & =\int\int\left\{ \frac{\partial\hat{\mathcal{L}_{P}}}{\partial\chi_{sp}^{\mu}}-\frac{D}{D\tau}\left[\frac{\partial\hat{\mathcal{L}_{P}}}{\partial\chi_{sp}^{\mu}}-\frac{1}{c^{2}}\left(\chi_{sp}^{\nu}\frac{\partial\hat{\mathcal{L}_{P}}}{\partial\dot{\chi}_{sp}^{\nu}}-\hat{\mathcal{L}_{P}}\right)\dot{\chi}_{sp\mu}\right]\right\} \delta\chi_{sp}^{\mu}d\tau d\Omega\nonumber \\
 &  & =\int\delta\chi_{sp}^{\mu}d\tau\int\left\{ \frac{\partial\hat{\mathcal{L}_{P}}}{\partial\chi_{sp}^{\mu}}-\frac{D}{D\tau}\left[\frac{\partial\hat{\mathcal{L}_{P}}}{\partial\dot{\chi}_{sp}^{\mu}}-\frac{1}{c^{2}}\left(\chi_{sp}^{\nu}\frac{\partial\hat{\mathcal{L}_{P}}}{\partial\dot{\chi}_{sp}^{\nu}}-\hat{\mathcal{L}_{P}}\right)\dot{\chi}_{sp\mu}\right]\right\} d\Omega.\label{eq:33}
\end{eqnarray}

The second and third terms on the right-hand side of Eq.\,\eqref{eq:30}
are actually 
\begin{equation}
\int\left\{ \frac{\partial\mathcal{L}}{\partial A_{\mu}}-\frac{D}{D\chi^{\nu}}\left[\frac{\partial\mathcal{L}}{\partial\left(\partial_{\nu}A_{\mu}\right)}\right]\right\} \delta A_{\mu}d\Omega.\label{eq:34}
\end{equation}
If we define

\begin{equation}
E_{A}^{\mu}=\frac{\partial\mathcal{L}}{\partial A_{\mu}}-\frac{D}{D\chi^{\nu}}\left[\frac{\partial\mathcal{L}}{\partial\left(\partial_{\nu}A_{\mu}\right)}\right],\label{eq:35}
\end{equation}
\begin{equation}
E_{\chi_{sp}\mu}=\frac{\partial\hat{\mathcal{L}_{P}}}{\partial\chi_{sp}^{\mu}}-\frac{D}{D\tau}\left[\frac{\partial\hat{\mathcal{L}_{P}}}{\partial\dot{\chi}_{sp}^{\mu}}-\frac{1}{c^{2}}\left(\chi_{sp}^{\nu}\frac{\partial\hat{\mathcal{L}_{P}}}{\partial\dot{\chi}_{sp}^{\nu}}-\hat{\mathcal{L}_{P}}\right)\dot{\chi}_{sp\mu}\right],\label{eq:36}
\end{equation}
and substitute Eqs.$\,$(\ref{eq:33})-(\ref{eq:36}) into Eq.$\,$(\ref{eq:30}),
then 
\begin{equation}
\delta\mathcal{A}=\int\left[\int E_{\chi_{sp}\text{\ensuremath{\mu}}}d\Omega\right]\delta\chi_{sp}^{\mu}d\tau+\int E_{A}^{\mu}\delta A_{\mu}d\Omega.\label{eq:37}
\end{equation}
Thus,

\begin{gather}
E_{A}^{\mu}=\frac{\partial\mathcal{L}}{\partial A_{\mu}}-\frac{D}{D\chi^{\nu}}\left[\frac{\partial\mathcal{L}}{\partial\left(\partial_{\nu}A_{\mu}\right)}\right]=0,\label{eq:38}\\
\int E_{\chi_{sp}\mu}d\Omega\equiv\int\left\{ \frac{\partial\hat{\mathcal{L}_{P}}}{\partial\chi_{sp}^{\mu}}-\frac{D}{D\tau}\left[\frac{\partial\hat{\mathcal{L}_{P}}}{\partial\dot{\chi}_{sp}^{\mu}}-\frac{1}{c^{2}}\left(\chi_{sp}^{\nu}\frac{\partial\hat{\mathcal{L}_{P}}}{\partial\dot{\chi}_{sp}^{\nu}}-\hat{\mathcal{L}_{P}}\right)\dot{\chi}_{sp\mu}\right]\right\} d\Omega=0,\label{eq:39}
\end{gather}
by the arbitrariness of $\delta\chi_{sp}^{\mu}$ and $\delta A_{\mu}$. 

Equation (\ref{eq:38}) is the EL equation of the 4-potential of the
electromagnetic field. Substituting Eq.$\,$(\ref{eq:24}) into Eq.$\,$(\ref{eq:38})
gives the Maxwell equation, 
\begin{equation}
\partial^{\nu}\left(\text{\ensuremath{\partial}}_{\nu}A_{\mu}-\text{\ensuremath{\partial}}_{\mu}A_{\nu}\right)=\frac{4\pi}{c}\sum_{s,p}q_{s}\int\dot{\chi}_{sp\mu}\delta\left(\chi_{sp}-\chi\right)ds,\label{eq:40}
\end{equation}
where $s$ is the length of the world-line, i.e., 
\begin{equation}
ds=cd\tau.
\end{equation}
The 4-current is 
\begin{equation}
J_{\mu}=\sum_{s,p}q_{s}\int\dot{\chi}_{sp\mu}\delta\left(\chi_{sp}-\chi\right)ds.
\end{equation}

Equation (\ref{eq:39}), will be called geometric sub-manifold EL
equation because it is defined only on the world-line after the integrating
over the space-time variable \cite{Qin2014}. If $\chi_{sp}$ were
a function of the entire space-time domain, then $E_{\chi_{sp}}$
would vanish everywhere, as in the case for 4-potential $A$. Generally,
we expect that $E_{\chi_{sp}}\neq0$.

We now derive an expression for $E_{\chi_{sp}}$ by substituting Eq.$\,$(\ref{eq:26})
into Eq.$\,$(\ref{eq:36}). For the first term in $E_{\chi_{sp}}$,
\begin{eqnarray}
\frac{\partial\hat{\mathcal{L}_{P}}}{\partial\chi_{sp}^{\mu}} & = & \left(-m_{s}c^{2}-\frac{q_{s}}{c}A_{\nu}\dot{\chi}_{sp}^{\nu}\right)\frac{\partial\delta_{2}}{\partial\chi_{sp}^{\mu}}\nonumber \\
 & = & \frac{D}{D\chi^{\mu}}\left[\left(m_{s}c^{2}+\frac{q_{s}}{c}A_{\nu}\dot{\chi}_{sp}^{\nu}\right)\delta_{2}\right]-\frac{q_{s}}{c}\frac{\partial A_{\nu}}{\partial\chi^{\mu}}\dot{\chi}_{sp}^{\nu}\delta_{2},\label{eq:43}
\end{eqnarray}
where $\delta_{2}\equiv\delta\left(\chi_{sp}-\chi\right)$. The second
term of $E_{\chi_{sp}}$ is given by 
\begin{eqnarray}
-\frac{D}{D\tau}\left(\frac{\partial\hat{\mathcal{L}_{P}}}{\partial\dot{\chi}_{sp}^{\mu}}\right) & = & \frac{q_{s}}{c}A_{\mu}\frac{D\delta_{2}}{D\tau}=-\frac{q_{s}}{c}A_{\mu}\dot{\chi}_{sp}^{\nu}\frac{\partial\delta_{2}}{\partial\chi^{\nu}}\nonumber \\
 & = & -\frac{D}{D\chi^{\nu}}\left[\frac{q_{s}}{c}A_{\mu}\dot{\chi}_{sp}^{\nu}\delta_{2}\right]+\frac{q_{s}}{c}\dot{\chi}_{sp}^{\nu}\frac{\partial A_{\mu}}{\partial\chi^{\nu}}\delta_{2}\label{eq:44}
\end{eqnarray}
The third and fourth terms came from the mass-shell condition, and
can be rewritten as 
\begin{eqnarray}
 &  & -\frac{D}{D\tau}\left(-\frac{1}{c^{2}}\dot{\chi}_{sp}^{\nu}\frac{\partial\hat{\mathcal{L}_{P}}}{\partial\dot{\chi}_{sp}^{\nu}}\dot{\chi}_{sp\mu}\right)\nonumber \\
 &  & =-\frac{q_{s}}{c^{3}}A_{\nu}\ddot{\chi}_{sp}^{\nu}\dot{\chi}_{sp\mu}\delta_{2}-\frac{q_{s}}{c^{3}}A_{\nu}\dot{\chi}_{sp}^{\nu}\ddot{\chi}_{sp\mu}\delta_{2}-\frac{q_{s}}{c^{3}}A_{\nu}\dot{\chi}_{sp}^{\nu}\dot{\chi}_{sp\mu}\frac{D\delta_{2}}{D\tau}\nonumber \\
 &  & =-\frac{q_{s}}{c^{3}}A_{\nu}\ddot{\chi}_{sp}^{\nu}\dot{\chi}_{sp\mu}\delta_{2}-\frac{q_{s}}{c^{3}}A_{\nu}\dot{\chi}_{sp}^{\nu}\ddot{\chi}_{sp\mu}\delta_{2}+\frac{D}{D\chi^{\sigma}}\left(\frac{q_{s}}{c^{3}}A_{\nu}\dot{\chi}_{sp}^{\nu}\dot{\chi}_{sp\mu}\dot{\chi}_{sp}^{\sigma}\delta_{2}\right)\nonumber \\
 &  & -\frac{q_{s}}{c^{3}}\dot{\chi}_{sp}^{\nu}\frac{\partial A_{\nu}}{\partial\chi^{\sigma}}\dot{\chi}_{sp}^{\sigma}\dot{\chi}_{sp\mu}\delta_{2},\label{eq:45}
\end{eqnarray}
and 
\begin{eqnarray}
 &  & -\frac{D}{D\tau}\left(\frac{1}{c^{2}}\hat{\mathcal{L}_{P}}\dot{\chi}_{sp\mu}\right)\nonumber \\
 &  & =-\frac{D}{D\chi^{\nu}}\left[\left(m_{s}+\frac{q_{s}}{c^{3}}A_{\sigma}\dot{\chi}_{sp}^{\sigma}\right)\dot{\chi}_{sp\mu}\dot{\chi}_{sp}^{\nu}\delta_{2}\right]+\frac{q_{s}}{c^{3}}\dot{\chi}_{sp}^{\nu}\frac{\partial A_{\sigma}}{\partial\chi^{\nu}}\dot{\chi}_{sp}^{\sigma}\dot{\chi}_{sp\mu}\delta_{2}\nonumber \\
 &  & +m_{s}\ddot{\chi}_{sp\mu}\delta_{2}+\frac{q_{s}}{c^{3}}A_{\sigma}\dot{\chi}_{sp}^{\sigma}\ddot{\chi}_{sp\mu}\delta_{2}+\frac{q_{s}}{c^{3}}A_{\sigma}\ddot{\chi}_{sp}^{\sigma}\dot{\chi}_{sp\mu}\delta_{2}.\label{eq:46}
\end{eqnarray}
Therefore, 
\begin{eqnarray}
E_{\chi_{sp}\text{\ensuremath{\mu}}} & = & \frac{D}{D\chi^{\nu}}\left\{ \left[\left(m_{s}c^{2}+\frac{q_{s}}{c}A_{\sigma}\dot{\chi}_{sp}^{\sigma}\right)\eta_{\mu}^{\;\text{\ensuremath{\nu}}}-\left(\frac{q_{s}}{c}A_{\mu}+m_{s}\dot{\chi}_{sp\mu}\right)\dot{\chi}_{sp}^{\nu}\right]\delta_{2}\right\} \nonumber \\
 & + & \left[m_{s}\ddot{\chi}_{sp\mu}-\frac{q_{s}}{c}\dot{\chi}_{sp}^{\nu}\left(\frac{\partial A_{\nu}}{\partial\chi^{\mu}}-\frac{\partial A_{\mu}}{\partial\chi^{\nu}}\right)\right]\delta_{2}.\label{eq:47}
\end{eqnarray}

Substituting Eq.$\,$(\ref{eq:47}) into the geometric sub-manifold
EL equation (\ref{eq:39}), we immediately obtain the equation of
motion for particles, 
\begin{equation}
m_{s}\ddot{\chi}_{sp\mu}=\frac{q_{s}}{c}\left(\frac{\partial A_{\nu}}{\partial\chi^{\mu}}-\frac{\partial A_{\mu}}{\partial\chi^{\nu}}\right)\dot{\chi}_{sp}^{\nu}.\label{eq:48}
\end{equation}
The right term in Eq.$\,$(\ref{eq:48}) is the 4-Lorenzen force on
particles. Then, $E_{\chi_{sp}}$ reduces to 
\begin{eqnarray}
E_{\chi_{sp}\text{\ensuremath{\mu}}} & = & \frac{\partial\hat{\mathcal{L}_{P}}}{\partial\chi_{sp}^{\mu}}-\frac{D}{D\tau}\left[\frac{\partial\hat{\mathcal{L}_{P}}}{\partial\dot{\chi}_{sp}^{\mu}}-\frac{1}{c^{2}}\left(\dot{\chi}_{sp}^{\nu}\frac{\partial\hat{\mathcal{L}_{P}}}{\partial\dot{\chi}_{sp}^{\nu}}-\hat{\mathcal{L}_{P}}\right)\dot{\chi}_{sp\mu}\right]\nonumber \\
 & = & \frac{D}{D\chi^{\nu}}\left\{ \left[\left(m_{s}c^{2}+\frac{q_{s}}{c}A_{\sigma}\dot{\chi}_{sp}^{\sigma}\right)\eta_{\mu}^{\;\nu}-\left(\frac{q_{s}}{c}A_{\mu}+m_{s}\dot{\chi}_{sp\mu}\right)\dot{\chi}_{sp}^{\nu}\right]\delta_{2}\right\} .\label{eq:49}
\end{eqnarray}
As expected, $E_{\chi_{sp}}\neq0$. We will refer to Eq.$\,$(\ref{eq:49})
as the geometric weak EL equation \cite{Qin2014}, which play a significant
role in subsequent analysis of the local conservation laws. The qualifier
``weak'' is used to indicate the fact that only the space-time integral
of $E_{\chi_{sp}}$ is zero. Unlike the non-relativistic case (see
Eq.$\,$(\ref{eq:27}) ), the geometric weak EL equation here is written
as a differential equation about $\hat{\mathcal{L}_{P}}$, instead
of the Lagrangian density. 

Another crucial step to systematically obtain conservation laws is
finding the symmetries. A symmetry group of action $\mathcal{A}$
is defined by a continuous transformation

\begin{equation}
\left(\chi,[\chi_{sp}],[\dot{\chi}_{sp}],A,\partial A\right)\longmapsto\left(\widetilde{\chi},[\widetilde{\chi}{}_{sp}],[\widetilde{\dot{\chi}}_{sp}],\widetilde{A},\widetilde{\partial A}\right)
\end{equation}
such that 
\begin{equation}
\int\mathcal{L}\left(\chi,[\chi_{sp}],[\dot{\chi}_{sp}],A,\partial A\right)d\Omega=\int\tilde{\mathcal{L}}\left(\widetilde{\chi},[\widetilde{\chi}{}_{sp}],[\widetilde{\dot{\chi}}_{sp}],\widetilde{A},\widetilde{\partial A}\right)d\Omega,\label{eq:50}
\end{equation}
where the indexes for physical quantities has been omitted to simplify
the notations. The symbol $\text{\ensuremath{\left[\beta\right]}}\left(\beta=\chi_{sp},\dot{\chi}_{sp},\widetilde{\chi}{}_{sp},\widetilde{\dot{\chi}}_{sp}\right)$
indicates that the geometric Lagrangian density $\mathcal{L}$ is
a functional of $\beta$. For the following transformation defined
by 
\begin{equation}
\left(\widetilde{\chi},\left[\widetilde{\chi}{}_{sp}\right],\left[\widetilde{\dot{\chi}}_{sp}\right],\widetilde{A},\widetilde{\partial A}\right)=\left(\chi+\epsilon X,\left[\chi_{sp}+\epsilon X\right],\left[\dot{\chi}_{sp}\right],A,\partial A\right),\quad\epsilon\in R
\end{equation}
where $X$ is a given constant 4-vector field, the condition (\ref{eq:50})
is satisfied because 
\begin{equation}
\mathcal{L}=\int\hat{\mathcal{L}_{P}}\left(\chi_{sp},\dot{\chi}_{sp},\chi,A\right)d\tau+\mathcal{L}_{F}=\int\hat{\mathcal{L}_{P}}\left(\chi+\epsilon X,\chi_{sp}+\epsilon X,\dot{\chi}_{sp},A\right)d\tau+\mathcal{L}_{F},\label{eq:52}
\end{equation}
with $\widetilde{A}(\widetilde{\chi})=A(\chi)=A(\widetilde{\chi}-\epsilon X)$
and $\widetilde{\partial A}(\widetilde{\chi})=\partial A(\chi)=\partial A(\widetilde{\chi}-\epsilon X)$
\cite{Qin2014}. Equation (\ref{eq:52}) should be transformed into
a partial differential equation before we apply it in deriving conservation
laws. The symmetry condition is 
\begin{equation}
\left[\frac{d\mathcal{L}}{d\epsilon}\right]_{\epsilon=0}\equiv0,\label{eq:53}
\end{equation}
which is equivalent to 
\begin{eqnarray}
0 & = & \int d\tau\left[\frac{d\left(\chi^{\mu}+\epsilon X^{\mu}\right)}{d\epsilon}\frac{\partial\hat{\mathcal{L}_{P}}}{\partial\left(\chi^{\mu}+\epsilon X^{\mu}\right)}+\sum_{s,p}\frac{d\left(\chi_{sp}^{\mu}+\epsilon X^{\mu}\right)}{d\epsilon}\frac{\partial\hat{\mathcal{L}_{P}}}{\partial\left(\chi_{sp}^{\mu}+\epsilon X^{\mu}\right)}\right]_{\epsilon=0}\nonumber \\
 & = & X^{\mu}\int\left(\frac{\partial\hat{\mathcal{L}_{P}}}{\partial\chi^{\mu}}+\sum_{s,p}\frac{\partial\hat{\mathcal{L}_{P}}}{\partial\chi_{sp}^{\mu}}\right)d\tau.
\end{eqnarray}
Therefore, 
\begin{equation}
\int\left[\frac{\partial\hat{\mathcal{L}_{P}}}{\partial\chi^{\mu}}+\sum_{s,p}\frac{\partial\hat{\mathcal{L}_{P}}}{\partial\chi_{sp}^{\mu}}\right]d\tau=0,\label{eq:55}
\end{equation}
or 
\begin{equation}
\frac{\partial\mathcal{L}}{\partial\chi^{\mu}}+\text{\ensuremath{\sum_{s,p}\int\frac{\partial\hat{\mathcal{L}_{P}}}{\partial\chi_{sp}^{\mu}}d\tau=0}}\label{eq:57}
\end{equation}
Due to the integral, Eq.\,(\ref{eq:57}) indicates that the corresponding
vector field of this symmetry for the particle-field systems is infinite
dimensional.

Next, we establish the connection between the symmetry of the particle-field
system, i.e., Eq.$\,$(\ref{eq:57}), and geometric conservation laws.
The first term in Eq.$\,$(\ref{eq:57}) is 
\begin{equation}
\text{\ensuremath{\frac{\partial\mathcal{L}}{\partial\chi^{\mu}}}=}\frac{D\mathcal{L}}{D\chi^{\mu}}-\frac{DA_{\nu}}{D\chi^{\mu}}\frac{\partial\mathcal{L}}{\partial A_{\nu}}-\frac{D\left(\partial_{\sigma}A_{\nu}\right)}{D\chi^{\mu}}\frac{\partial\mathcal{L}}{\partial\left(\partial_{\sigma}A_{\nu}\right)}=\frac{D}{D\chi^{\nu}}\left[\mathcal{L}\eta_{\mu}^{\;\nu}-\partial_{\mu}A_{\sigma}\frac{\partial\mathcal{L}}{\partial\left(\partial_{\nu}A_{\sigma}\right)}\right],\label{eq:58}
\end{equation}
where use is made of the EL equation, i.e., Eq.$\,$(\ref{eq:38}).
For second term in Eq.$\,$(\ref{eq:57}), we use the geometric weak
EL equation (\ref{eq:49}) to obtain 
\begin{eqnarray}
\int_{p_{1}}^{p_{2}}\frac{\partial\hat{\mathcal{L}_{P}}}{\partial\chi_{sp}^{\mu}}d\tau & = & \frac{D}{D\chi^{\nu}}\left\{ \int\left[\left(m_{s}c^{2}+\frac{q_{s}}{c}A_{\sigma}\dot{\chi}^{\sigma}\right)\eta_{\mu}^{\;\nu}\delta_{2}-\left(\frac{q_{s}}{c}A_{\mu}+m_{s}\dot{\chi}_{sp\mu}\right)\dot{\chi}_{sp}^{\nu}\delta_{2}\right]d\tau\right\} \nonumber \\
 & + & \left[\frac{\partial\hat{\mathcal{L}_{P}}}{\partial\dot{\chi}_{sp}^{\mu}}-\frac{1}{c^{2}}\left(\chi_{sp}^{\nu}\frac{\partial\hat{\mathcal{L}_{P}}}{\partial\dot{\chi}_{sp}^{\nu}}-\hat{\mathcal{L}_{P}}\right)\dot{\chi}_{sp\mu}\right]_{p_{1}}^{p_{2}}.\label{eq:60}
\end{eqnarray}
The boundaries, i.e., $p_{1}$ and $p_{2}$ of the integral of Eq.$\,$(\ref{eq:60})
should be extended to infinity, which will make the last term vanishes
because of the existence of the Dirac delta function in $\hat{\mathcal{L}_{P}}.$

Bringing Eqs.$\,$(\ref{eq:58}), (\ref{eq:60}) and \ref{eq:24}
into Eq.$\,$(\ref{eq:57}), we obtain

\begin{equation}
\frac{D}{D\chi^{\nu}}\left[\sum_{s,p}m_{s}\int\dot{\chi}_{sp\mu}\dot{\chi}_{sp}^{\nu}\delta_{2}ds-\frac{1}{2\pi}\partial_{\mu}A_{\sigma}\partial^{[\nu}A^{\sigma]}-\frac{1}{4\pi}A_{\mu}\partial^{\sigma}F_{\sigma}^{\;\nu}+\frac{1}{16\pi}F_{\sigma\rho}F^{\sigma\rho}\eta_{\mu}^{\;\nu}\right]=0,\label{eq:62}
\end{equation}
which is equivalent to 
\begin{equation}
\frac{D}{D\chi^{\nu}}\left\{ \sum_{s,p}m_{s}\int\dot{\chi}_{sp\mu}\dot{\chi}_{sp}^{\nu}\delta_{2}d\tau+\frac{1}{4\pi}\left(-F_{\mu}^{\;\sigma}F_{\;\sigma}^{\nu}+\frac{1}{4}\eta_{\mu}^{\;\nu}F_{\sigma\rho}F^{\sigma\rho}\right)-\frac{1}{4\pi}\left[\partial_{\sigma}\left(A_{\mu}F^{\nu\sigma}\right)\right]\right\} =0,\label{eq:64}
\end{equation}
with the identity

\begin{equation}
-\frac{1}{2\pi}\partial_{\mu}A_{\sigma}\partial^{[\nu}A^{\sigma]}-\frac{1}{4\pi}A_{\mu}\partial^{\sigma}F_{\sigma}^{\;\nu}=-\frac{1}{4\pi}F_{\mu\sigma}F^{\nu\sigma}-\frac{1}{4\pi}\partial_{\sigma}\left(A_{\mu}F^{\nu\sigma}\right).
\end{equation}
The last term in Eq.$\,$(\ref{eq:64}) is zero because 
\begin{equation}
-\frac{1}{4\pi}\frac{D}{D\chi^{\nu}}\partial_{\sigma}\left(A_{\mu}F^{\nu\sigma}\right)=-\frac{1}{4\pi}\frac{D}{D\chi^{(\nu}}\frac{D}{D\chi^{\sigma)}}\left(A_{\mu}F^{[\nu\sigma]}\right)\equiv0,
\end{equation}
where $(\nu\;\mu)$ is the total symmetrization of the indexes of
$\nu$ and $\mu$.  Finally, we arrive at the geometric, or manifestly
covariant, energy-momentum conservation laws 
\begin{equation}
\frac{\partial}{\partial\chi^{\nu}}\left\{ \sum_{s,p}m_{s}\int\dot{\chi}_{s}^{\mu}\dot{\chi}_{sp}^{\nu}\delta_{2}ds+\frac{1}{4\pi}\left(-F^{\text{\ensuremath{\mu\sigma}}}F_{\;\sigma}^{\nu}+\frac{1}{4}\eta^{\mu\nu}F_{\sigma\rho}F^{\sigma\rho}\right)\right\} =0,\label{eq:67}
\end{equation}
where we regard the field as defined on $\chi$, and $\partial/\partial\chi^{\nu}\equiv D/D\chi^{\nu}$.
In terms of energy-momentum tensor, Eq.\,(\ref{eq:67}) is 
\begin{equation}
\partial_{\nu}T^{\mu\nu}=0,
\end{equation}
where 
\begin{gather}
T^{\mu\nu}=T_{P}^{\mu\nu}+T_{F}^{\mu\nu},\\
T_{P}^{\mu\nu}=\sum_{s,p}m_{s}\int\dot{\chi}_{sp}^{\mu}\dot{\chi}_{sp}^{\nu}\delta_{2}ds,\label{eq:68}\\
T_{F}^{\mu\nu}=\frac{1}{4\pi}\left(-F^{\text{\ensuremath{\mu\sigma}}}F_{\;\sigma}^{\nu}+\frac{1}{4}\eta^{\mu\nu}F_{\sigma\rho}F^{\sigma\rho}\right).\label{eq:69}
\end{gather}
Here, $T_{P}^{\mu\nu}$, $T_{F}^{\mu\nu}$ and $T^{\mu\nu}$ are the
energy-momentum tensors of the particles, electromagnetic field and
the particle-field systems written in the geometric form, respectively,
and it's easy to check that all of these tensors are symmetric. The
energy-momentum tensor $T_{F}^{\mu\nu}$ for electromagnetic field
described by Eq.$\,$(\ref{eq:69}) is well-known. However, to the
best of our knowledge, the geometric energy-momentum tensor for particles
given in Eq.\,\eqref{eq:68} has not been derived previously. Instead,
the energy-momentum tensor for particle is typically written in the
existing literature \cite{Landau1975} as

\begin{equation}
T_{P}^{\mu\nu}=\sum_{s,p}m_{s}\delta\left(\mathbf{X}_{sp}(t)-\boldsymbol{x}\right)\dot{\chi}_{sp}^{\mu}(t)\dot{\chi}_{sp}^{\nu}(t)\frac{d\tau(t)}{dt},\label{eq:71}
\end{equation}
where a Lorentzian coordinate system $\{t,\boldsymbol{x}\}$ is chosen.
Obviously, Eq.$\,$(\ref{eq:71}) is not a geometric form as Eq.\,\eqref{eq:68}.
We can prove that Eq.$\,$(\ref{eq:68}) recovers Eq.$\,$(\ref{eq:71})
when a coordinate system is chosen. For this purpose, we first show
that 
\begin{equation}
\int\delta\left(\chi_{sp}-\chi\right)g\left(\chi_{sp}\right)ds=\gamma_{sp}^{-1}(t)\delta\left(\mathbf{X}_{sp}(t)-\boldsymbol{x}\right)g\left(\chi_{sp}\left(t\right)\right),\label{eq:72}
\end{equation}
where $\gamma_{sp}^{-1}(t)\equiv\sqrt{1-\dot{\mathbf{x}}_{sp}^{2}\left(t\right)/c^{2}}$,
and $g\left(\chi_{sp}\right)$ is an arbitrary field. The left-hand
side of Eq.$\,$(\ref{eq:72}) is 
\begin{eqnarray}
 &  & \int\delta\left(\chi_{sp}-\chi\right)g\left(\chi_{sp}\right)ds\nonumber \\
 &  & \equiv\int\delta\left[\chi_{sp}\left(s_{sp}\right)-\chi\right]g\left[\chi_{sp}\left(s_{sp}\right)\right]ds_{sp}\\
 &  & =\int\delta\left[\chi_{sp}\left(s_{sp}\left(t_{sp}\right)\right)-\chi\right]g\left[\chi_{sp}\left(s_{sp}\left(t_{sp}\right)\right)\right]\frac{ds_{sp}}{dt_{sp}}dt_{sp}\nonumber \\
 &  & =\int\delta\left[c\left(t_{sp}-t\right)\right]\delta\left[\mathbf{X}_{sp}\left(t_{sp}\right)-\boldsymbol{x}\right]g\left[\chi_{sp}\left(t_{sp}\right)\right]\frac{cd\tau_{sp}\left(t_{sp}\right)}{dt_{sp}}dt_{sp},\label{eq:73}
\end{eqnarray}
where $t_{sp}$ is the time parameter for each world-line, and $\boldsymbol{x}_{sp}\left(t_{sp}\right)$
is the space position for $sp$-particle at time $t_{sp}$. Because
\begin{equation}
\frac{d\tau_{sp}\left(t_{sp}\right)}{dt_{sp}}=\gamma_{sp}^{-1}(t_{sp}),\label{eq:74}
\end{equation}
and 
\begin{equation}
\delta\left[c\left(t_{sp}-t\right)\right]=\frac{1}{c}\delta\left(t_{sp}-t\right),\label{eq:75}
\end{equation}
we have

\begin{align}
\int\delta\left(\chi_{sp}-\chi\right)g\left(\chi_{sp}\right)ds & =\int\gamma_{sp}^{-1}(t_{sp})\delta\left[\mathbf{X}_{sp}\left(t_{sp}\right)-\boldsymbol{x}\right]g\left[\chi_{sp}\left(t_{sp}\right)\right]\delta\left(t_{sp}-t\right)dt_{sp}\nonumber \\
 & =\gamma_{sp}^{-1}(t)\delta\left[\mathbf{X}_{sp}\left(t\right)-\boldsymbol{x}\right]g\left[\chi_{sp}\left(t\right)\right],
\end{align}
which is Eq.$\,$(\ref{eq:72}). If we take $g\left(\chi_{sp}\right)=m_{s}\dot{\chi}_{sp}^{\mu}\dot{\chi}_{sp}^{\nu},$
then the geometric energy-momentum tensor for particles is 
\begin{align}
T_{P}^{\mu\nu} & =\sum_{s,p}m_{s}\int\delta\left(\chi_{sp}-\chi\right)\dot{\chi}_{sp}^{\mu}\dot{\chi}_{sp}^{\nu}ds=m_{s}\gamma_{sp}^{-1}(t)\delta\left[\mathbf{X}_{sp}\left(t\right)-\boldsymbol{x}\right]\dot{\chi}_{sp}^{\mu}\left(t\right)\dot{\chi}_{sp}^{\nu}\left(t\right)\nonumber \\
 & =\sum_{s,p}m_{s}\delta\left[\mathbf{X}_{sp}\left(t\right)-\boldsymbol{x}\right]\dot{\chi}_{sp}^{\mu}\left(t\right)\dot{\chi}_{sp}^{\nu}\left(t\right)\frac{d\tau_{sp}\left(t\right)}{dt}.\label{eq:80}
\end{align}
This confirms that the geometric energy-momentum tensor for particles
recovers the non-geometric form in a chosen coordinate system.

\section{Conclusions\label{sec:Conclusions}}

In this paper, we have developed a manifestly covariant, or geometric,
field theory for the relativistic classical particle-field systems
often encountered in plasma physics, accelerator physics, and astrophysics.
The connection between space-time symmetry and energy-momentum conservation
laws is demonstrated geometrically. In our theoretical formalism,
space and time are treated with equal footing, i.e., space-time is
treated as one identity without choosing a coordinate system. This
is different from existing field theories where it is necessary to
split space and time coordinates at certain stage, and thus the manifestly
covariant property is lost. 

There are several unique features in the geometric field theory developed.
The first is the mass-shell condition, which induces two new terms
in the geometric EL equation (\ref{eq:19}) for particles. The geometric
Lagrangian density of particle-field systems is a functional of particles'
world-lines (see equation (\ref{eq:24})), which makes the symmetry
vector field of the systems lies on the infinitive dimensional space
(see Eq.$\,$(\ref{eq:60})). Another feature of the theory is that
particles and fields reside on different manifolds. The domain of
the particle field can be the proper time or any other parameterization
for the world-lines, and the electromagnetic field, on the other hand,
are defined on space-time. In order to establish geometrically the
connection between symmetries and energy-momentum conservations, a
geometric weak EL equation (\ref{eq:48}) for particles is derived.
Combining the EL equation (\ref{eq:38}) for field and the geometric
weak EL equation (\ref{eq:48}) for particles, symmetries and conservation
laws could be established geometrically. Using the theory, we derived
for the first time a geometric expression for the energy-momentum
tensor for particles in Eq.$\,$(\ref{eq:68}), which recovers the
non-geometric form in the existing literature \cite{Landau1975} for
a chosen coordinate system.

In the present study, we make use of proper time. We note that different
particles have different proper time, which is not synchronized in
the laboratory frame. This brings difficulties if one would like to
use proper time for particle-in-cell (PIC) simulations. However, proper
time can be beneficial in certain situations. For example, proper
time has been utilized to construct explicit symplectic integrators
for relativistic dynamics of charged particles \cite{Wang2016,Zhou2017}.
As a matter of fact, the geometric field theory for classical particle-field
systems in the present study can only be established with the help
of proper time. As for the specific application of proper time in
PIC simulations, more investigation is needed. The very facts that
proper time can be used to construct explicit symplectic integrators
and that it is essential in establishing the geometric field theory
for classical particle-field systems suggest that proper time could
play a role in developing advanced PIC algorithms \cite{Squire2012,Xiao2015,Qin2016}.
For example, we can investigate the possibility of using different
proper time-steps in the symplectic integrators for different particles
such that they are synchronized in the laboratory frame. This topic
will be explored in future study. 
\begin{acknowledgments}
This research is supported by National Magnetic Confinement Fusion
Energy Research Project (2015GB111003, 2014GB124005), National Natural
Science Foundation of China (NSFC-11575185, 11575186, 11305171), JSPS-NRF-NSFC
A3 Foresight Program (NSFC-11261140328), Key Research Program of Frontier
Sciences CAS (QYZDB-SSW-SYS004), the Geo-Algorithmic Plasma Simulator
(GAPS) Project, and National Magnetic Confinement Fusion Energy Research
Project ( 2013GB111002B ). 
\end{acknowledgments}

\bibliography{GeometricWEL}

\end{document}